\documentstyle[ApJ,Times]{article}

\begin{document}

\def\simg{\mathrel{%
      \rlap{\raise 0.511ex \hbox{$>$}}{\lower 0.511ex \hbox{$\sim$}}}}
\def\siml{\mathrel{%
      \rlap{\raise 0.511ex \hbox{$<$}}{\lower 0.511ex \hbox{$\sim$}}}}
\def\Mesz{M\'esz\'aros~}
\def\ie{i.e$.$~} \def\eg{e.g$.$~} \def\etal{et al$.$~}
\def\epsel{\varepsilon_{el}} \def\epsmag{\varepsilon_{mag}} 
\def\epsmg1{\varepsilon_{mag,-1}} \def\eq{eq$.$~}

\begin{center}
\title{\sc Hydrodynamical Simulations of Gamma Ray-Bursts \\
        from Internal Shocks in Relativistic Fireballs}

\author{A. Panaitescu and P. \Mesz}
\affil{Department of Astronomy \& Astrophysics,
    Pennsylvania State University, University Park, PA 16802}
\end{center}

\begin{abstract}
  We simulate the dynamics of the interaction between relativistically 
  expanding spherically symmetric shells using a 1-dimensional hydrodynamic
  code and calculate spectra and light-curves arising from such collisions
  by integrating the synchrotron and inverse Compton emission of the
  shocked gas. The numerical results reflect the most important features
  observed in Gamma-Ray Bursts: the spectrum exhibits a progressive
  softening (its break energy decaying exponentially with the 50--300 keV
  photon fluence), and the pulses that form the burst appear narrower 
  in higher energy bands. 
  Analytical results for the most important physical parameters of the
  burst are obtained by solving the shock jump conditions for a pair of 
  interacting shells, in the case when both the forward and reverse shocks
  are relativistic.
\end{abstract}


\section{Introduction}

We consider Gamma-Ray Bursts (GRBs) arising from multiple internal
collisions between relativistic shells of cold ejecta expanding into 
a vacuum. Under typical conditions, such collisions take place well before 
the decelerating effect caused by an external medium becomes important.
The dynamics and energetics of the unsteady outflow leading to internal
shocks in an ejecta released, \eg in a compact merger or collapse event,
were outlined in Rees \& \Mesz (1994). In this model, the physical
conditions determining the energy deposition in the ejecta are not
steady during the entire event, resulting in a non-uniform distribution
of the bulk Lorentz factors within the ejecta. Faster shells of material
catch up with slower ones, leading to the formation of two shocks
in each pair of interacting shells. The slower shell is swept up by
a forward shock that accelerates it, while the faster ejecta are
decelerated by a reverse shock. Both shocks heat the ejecta, give rise 
to a turbulent magnetic field and accelerate electrons to a power-law
distribution. The shocked fluid cools through synchrotron and inverse 
Compton emission and through adiabatic losses, due to the radial expansion 
of the ejecta. The radiation received by the observer comes from a small 
part of the ejecta, that moves almost toward the observer, due to the 
high Lorentz factor of the emitting fluid. Consequently, the burst seen 
by the observer is substantially blue-shifted and much shorter than as
seen in the co-moving frame of the ejecta.   

The average spectra arising from internals shocks in such unsteady winds 
have been calculated analytically by Papathanassiou \& \Mesz (1996) for wide 
ranges of model parameters (magnetic field and electron acceleration 
efficiency, wind duration and wind variability timescale, average Lorentz 
factor of the ejecta). Their results show that burst spectra can extend over 
many orders of magnitude, from eV to TeV. The spectra calculated by Pilla \& 
Loeb (1998) show a similar wide range of photon energies, and a time evolution 
characterized by a softening of the spectrum, determined by the electron 
synchrotron and inverse Compton cooling in the early phase of the burst, and 
dominated by the pair-cascade process at later times. For reasonable values 
of the outflow Lorentz factors and the
shock radius, the depletion of high energy photons by pair creation was
found to alter the burst spectra at energies higher than the BATSE window. 
The two radiating processes which cool the electrons have been discussed 
by Sari \& Piran (1997), who derived constraints on model parameters from the
required efficiency, observed peak of the spectrum and ratio of the variability
timescale to the burst duration and from the condition of optical thinness 
to pair creation. A detailed parameter search of the properties of
time-integrated internal (as well as external) shocks provides constraints
on the values of parameters which lead to bursts in the BATSE window
(Papathanassiou \& \Mesz 1998).
Using a code based on the kinematics of the interaction 
between many shells, Kobayashi, Piran and Sari (1997) calculated the
efficiency with which shell collisions convert their bulk kinetic energy 
into internal energy and, assuming that this is radiated away, obtained a set 
of bolometric light-curves. In a more elaborate treatment, Daigne \& 
Mochkovitch (1998) studied the temporal and spectral features of the 
synchrotron emission 
from a population of electrons altered by the power-law injection at shocks. 
They found a good
agreement between most of their results and the features shown by real bursts: 
duration-hardness anti-correlation (\eg Kouveliotou \etal 1993), appropriate 
temporal asymmetry of pulses (Norris \etal 1996), spectra well fit by the 
Band function (Band \etal 1993), spectral hardening before a count rate increase
(\eg Bhat \etal 1995), pulse duration -- energy dependence consistent with 
observations (\eg Norris \etal 1996), and an exponential dependence of pulses 
peak energy on the photon fluence (Liang \& Kargatis 1996). 
Other features of the real bursts, \eg the general softening of 
the spectrum (Ford \etal 1995), are not well reproduced by the results 
presented by Daigne \& Mochkovitch (1998), probably due to the fact that they
did not take into account the effects arising from the geometrical curvature
of the emitting source or the radiative (synchrotron and inverse Compton) 
cooling of the electrons. 
 
In this work we investigate some of the most important properties 
of internal shock bursts, using a 1-dimensional hydrodynamic code suitable
for simulating the interaction between perfect fluids, and calculate the 
light-curve and spectrum of the radiation in the BATSE range emitted by 
shock-accelerated relativistic electrons, assuming spherical symmetry. In the 
framework of numerical hydrodynamic calculations we can follow in detail the
post-shock time evolution of the electrons due to radiative and adiabatic 
losses in each cell of the fluid, and integrate the synchrotron output to 
determine the photon field that is up-scattered by the electrons.
We then integrate the emission of the shocked fluid taking into account the
dependence of the Doppler boosting on the angle between the radial direction 
of outflow and the direction toward the observer, and the effect of the
shells' curvature on the photon arrival time. The bursts obtained
with this time-dependent radiation and hydrodynamic treatment of the shock
evolution show features that cannot be accounted for by kinematic treatments, 
including a softening of the spectrum and other correlations typical of 
the observed bursts.

\vspace*{3mm}
\section{Analytical Treatment of the Two-Shell Interaction}

Before proceeding to a numerical hydrodynamical treatment, it is useful to
start with some preliminary analytical insights. The physical conditions 
inside the shocked shells and the properties of the emitted radiation can
be estimated by first calculating the Lorentz factor of the shock that sweeps	
up a shell during the interaction with another relativistic shell, in the frame 
of the yet un-shocked part of the shell. This can be done using
the shock equations given by Blandford \& McKee (1976) and the fact that
the flow velocity and pressure in the shocked fluid has the same value
on both sides of the contact discontinuity that separates the interacting
shells. In the following analytic treatment, the internal pressure of
the un-shocked fluid is neglected, therefore it applies only to a pair of 
cold shells. The effect of the radial expansion during the shells' collision is also
neglected, therefore the following results are accurate only for shells
so thin that they do not expand significantly before the shocks sweep them.
If the two shells expand prior to their interaction in the way prescribed
by \Mesz, Laguna \& Rees (1993) (see below) then, during the time it takes 
the shocks to sweep up the ejecta, the radius of shells roughly doubles, 
therefore the radial expansion cannot be neglected and the following analytic
results should be regarded only as approximative.
 
 The inner shell has a co-moving frame density $\rho_i$ and
moves at a Lorentz factor $\Gamma_i$, while the outer shell has a co-moving
density $\rho_o$ and moves at a Lorentz factor $\Gamma_o$, lower 
than that of the inner shell. The collision of the two shells generates 
two shocks that compress and heat them.  If $\Gamma_{sh}$ denotes the Lorentz 
factor of the shocked fluid in the lab-frame, then the Lorentz factors 
$\Gamma'_{sh,i}$ and $\Gamma'_{sh,o}$ of this fluid in the frame of the 
yet un-shocked parts of the inner and, respectively, outer shell are:
\begin{equation}
 \Gamma'_{sh,i/o}=\frac{1}{2} \left(\frac{\Gamma_{i/o}}{\Gamma_{sh}} 
                    +\frac{\Gamma_{sh}}{\Gamma_{i/o}}\right) \;,
\end{equation}
or equivalently:
\begin{equation}
 \Gamma'_{sh,i}=\frac{x^2+g^2}{2gx} \,, \qquad  
 \Gamma'_{sh,o}=\frac{g^2x^2+1}{2gx} \;,
\label{Gprim}
\end{equation}
where 
\begin{equation}
 x \equiv \frac{\Gamma_{sh}}{\sqrt{\Gamma_{i}\Gamma_o}}\,, \quad 
 g \equiv \sqrt{\frac{\Gamma_i}{\Gamma_o}} > 1 \,, \quad
 \frac{1}{g} < x < g \;.
\label{def}
\end{equation}
The pressure of the shocked fluid around the contact discontinuity can be calculated for
each shock:
\begin{equation}
 (p_{sh})_{i/o} = (\Gamma'_{sh,i/o}-1) (\hat{\gamma}\Gamma'_{sh,i/o}+1)
              \rho_{i/o}c^2 \;,
\label{P}
\end{equation}
where $\hat{\gamma}$ is the adiabatic index of the heated fluid, assumed to be
the same on both sides of the contact discontinuity.
Using the equality of these pressures and equation (\ref{Gprim}), one obtains
a quartic equation for the lab-frame Lorentz factor of the shocked fluid:
\begin{displaymath}
 \hat{\gamma}(g^4-Y)x^4 + 2(\hat{\gamma}-1)g(Y-g^2)x^3 + 2(2-\hat{\gamma})
  g^2(Y-1)x^2 +
\end{displaymath} 
\begin{equation}
 2(\hat{\gamma}-1)g(g^2Y-1)x + \hat{\gamma}(1-g^4Y) = 0 \;,
\label{xeq}
\end{equation}
where $Y \equiv \rho_i/\rho_o$.

 Generally, equation (\ref{xeq}) can be solved only numerically. Figure 1
shows the Lorentz factors $\Gamma_{sh,i}$ and $\Gamma_{sh,o}$ of the shocked 
fluid as functions of the ratio
of the pre-shock co-moving densities for three values of the ratio of the
pre-shock lab frame Lorentz factors. For given parameter $g$, the $\Gamma'_{sh,i}$
and  $\Gamma'_{sh,o}$ curves are symmetric relative to the ordinate, \ie
$\Gamma'_{sh,i}(Y)=\Gamma'_{sh,o}(Y^{-1})$, as it can be shown using
equations (\ref{Gprim}) and (\ref{xeq}). It can be seen that, unless
$\Gamma_{i}/\Gamma_o \simg 5$ and either $Y \simg 10^2$ or $Y^{-1} \simg 10^2$, 
then $\Gamma'_{sh,i/o} < 2$, which implies that the two shocks that sweep the shells 
(the reverse shock in the inner shell and the forward shock in the outer one) 
are only mildly relativistic. Figure 1
also shows the efficiency $\epsilon$ at which the two shocks convert the shells'
kinetic energy into internal, defined as the ratio between the lab-frame
internal energy of the shocked gas and the kinetic energy that this gas had 
before it was swept up by one of the shocks. For given parameters $g$ and $Y$, 
the efficiency is constant as long as there is a shock sweeping up each shell. 
The efficiency of kinetic to internal energy conversion changes after one of 
the shells has been entirely swept up by one of the shocks; in this case one 
can calculate the overall efficiency of the interaction from conservation of 
momentum and energy. The result is:
\begin{equation}
 \epsilon = 1 - \left[1+ \frac{\mu}{G} \left(\frac{G-1}{\mu+1}\right)^2 
       \right]^{-1/2} \;,
\end{equation}
where $\mu=M_i/M_o$ is the ratio of the shells' masses and 
$G=g^2=\Gamma_i/\Gamma_o$ is the ratio of their Lorentz factors.

 Analytic solutions of equation (\ref{xeq}) can be obtained in the
particular case $Y=1$ or if one makes the assumption that the
two shocks are either quasi-newtonian or relativistic, which requires
that $\Gamma_i \simg \Gamma_o$ or $\Gamma_i \gg \Gamma_o$, 
respectively. These solutions are:
\begin{equation}
 Y=1: x=1, \quad G-1 \ll 1: x=\frac{gy+1}{g+y}, 
           \quad G \gg 1: x=\frac{Gy-1}{G-y} \,,
\end{equation}
where $y = \sqrt{Y} = \sqrt{\rho_i/\rho_o}$. 
Once the shock Lorentz factors are known, one can calculate
the co-moving internal and rest mass energy density of the shocked fluid,
the turbulent magnetic field $B$, and the minimum electron Lorentz factor 
$\gamma_m$ of the power-law distribution in each shocked shell, from where 
all the important characteristics of the
synchrotron and inverse Compton emission can be derived. 
Some analytical results in this direction are presented in Appendix A, only
for the case of relativistic shocks ($\Gamma_i \gg \Gamma_o$), as this
case leads to higher burst efficiency and, thus, is more likely to be
encountered in the bursts we observe.
For definiteness we consider here the case of two shells that have not 
undergone any previous collisions, in which case the two unknown pre-shock
co-moving densities $\rho_i$ and $\rho_o$ can be correlated with
basic burst parameters. The shells expand prior to the their interaction 
as described by \Mesz \etal (1993): they go through an acceleration
phase, coast at constant Lorentz factor, and later start expanding, their 
lab-frame thicknesses evolving as $r/\Gamma_{i/o}^2$, where $r$ is the 
radial coordinate. If the slower shell was released a time $t_v$ before the 
faster one, then the collision takes place at the ``interaction radius"
\begin{equation}
 r_{int} = 2 \Gamma_i^2 \frac{ct_v}{G^2-1}\;,
\label{rint}
\end{equation}
where the co-moving densities are
\begin{equation}
 \rho_{i/o}c^2 = \frac{E_{i/o}}{4\pi r_{int}^3}\;,
\label{rho}
\end{equation}
$E_i$ and $E_o$ being the kinetic energies of the two shells, therefore 
$Y=E_i/E_o$.

 Equations (\ref{syR}) and (\ref{icR}) show that 
the burst spectrum depends strongly on the Lorentz factors of the two shells,
the ratio of their energies, the burst variability timescale, and the electron 
acceleration efficiency. The dependence on the Lorentz factors and variability
timescale arises mainly from the fact that we considered the interaction between 
shells that have propagated unperturbed until their collision; the results would 
be different if one or both shells interacted before with other shells, in which 
case equation (\ref{rho}) does not hold. The spectrum dependence on the ratios 
of the shells' Lorentz factors and of their energies comes from the hydrodynamics 
of the interaction.

\vspace*{3mm}
\section{Description of the Numerical Code}

The numerical code that we developed contains two major parts: one which 
simulates the hydrodynamics of the interaction between two relativistically 
expanding fluids, described by Wen, Panaitescu \& Laguna (1997), and one which 
calculates the emission of radiation from the shocked gases, through 
synchrotron and inverse 
Compton processes, and computes the observed spectrum and photon/energy 
light-curves, described by Panaitescu \& \Mesz (1998a,b). The most important 
features of our calculation of the burst emission are listed below.

\vspace*{2mm}
 {\bf Electron distribution and magnetic field intensity.}
The intensity $B$ of the turbulent magnetic field is parameterized by
the fraction $\epsmag$ of the internal energy that is stored in the magnetic
field.  An electron power-law distribution (of exponent $-p$) is initialized in a grid 
cell containing shocked fluid when that cell is added to the shocked structure. 
The addition of a new ``shocked" cell is done by the Glimm method that we 
use to simulate the propagation of the shocks, when the mass of the pre-shock 
fluid swept-up since the last added cell reaches the mass corresponding to the
cell volume and post-shock density determined by the shock jump equations.  
The minimum electron random Lorentz factor $\gamma_m$ of the power-law 
distribution is set by the energy given to 
leptons after shock heating, taken as a constant fraction $\epsel$ of the total 
internal energy of the newly shocked fluid, and by the fraction $\zeta$ of 
electrons that are picked up by shock acceleration. The electrons are considered
decoupled from protons and magnetic fields after their initial acceleration
(\ie they are not re-energized), and lose energy through emission of synchrotron 
and inverse Compton radiation and through adiabatic cooling.

\vspace*{2mm}
 {\bf Radiative losses.}
Given the local value of the turbulent magnetic field $B$
and the evolving electron distribution in each shocked cell, at some 
lab-frame time $t$, we calculate the synchrotron losses and integrate the 
emitted radiation over the entire volume of the shocked fluid and over the 
electron distribution, to calculate the synchrotron radiation energy density
at each point in the shocked fluid, necessary for the computation of the 
inverse Compton losses. The approximations used for a faster numerical 
calculation of the synchrotron spectrum and inverse Compton losses are given
in Appendix B. Also to reduce the computational effort, we do not use the 
full shape of the synchrotron spectrum when calculating the inverse Compton 
losses; instead we approximate as monochromatic the synchrotron radiation to be 
up-scattered, at an intensity-weighted frequency.
Further, the spectrum of the up-scattered radiation is 
approximated as monochromatic, at the peak frequency of the inverse Compton 
spectrum corresponding to given electron Lorentz factor and to the 
intensity-averaged synchrotron frequency. Thus we expect that the up-scattered 
radiation is correctly calculated at frequencies that are not too close to the 
limits of the inverse Compton spectrum. We checked the correctness of this 
supposition by calculating burst light-curves using the full shape of the 
inverse Compton spectrum (which leads to substantially longer runs) and by 
comparing them with those obtained using the monochromaticity approximations. 
It is worth stressing here that, for an accurate treatment of the inverse 
Compton losses, one has to resort to numerics in order to take into account 
the relativistic beaming of the local synchrotron output, due to the 
relative motion of the cells where the photons are generated and up-scattered. 

\vspace*{2mm}
 {\bf Light-curves and spectra.}
We integrate the synchrotron and inverse Compton emission over the electron 
distribution, the volume of the entire shocked fluid, and the evolution of the 
interacting shells, to calculate the observed light-curves and 
instantaneous/brightness-averaged spectra, taking into 
account the beaming, Doppler frequency shift and time contraction due 
to the relativistic motion of the radiating fluid. 
The relativistic effects are dependent on the angle between the radial 
direction of outflow and the direction toward the observer. Therefore, 
assuming that the shells are spherically symmetric at least within the cone 
of half-angle $\Gamma^{-1}$ visible to the observer, the integral over volume 
is a double one (over shocked cells and over the angle relative to 
the line of sight toward the observer), which makes the burst spectrum and 
light-curve to be a quadruple integral.

\vspace*{3mm}
\section{Numerical Spectra and Light-Curves}
  
There are a number of model parameters which affect the burst light-curve 
in the 100 keV--300 keV range spectrum. To simplify things,
we consider here the interaction between two spherically symmetric shells that 
have fixed $E_i=10^{53}$ ergs, $E_o=2\times 10^{52}$ ergs, $\Gamma_i=100$
and $\Gamma_o=50$. The initial kinetic energy of the shells can be much lower 
if they are emitted within a relatively narrow cone; these values were chosen so
that, for a burst located at redshift $z=1$ and having the low efficiency 
corresponding to ($G=2,Y=5$), the fluence in the BATSE window is above 
$10^{-7}\,{\rm erg\, cm^{-2}}$. To maximize the brightness of the burst we 
assume that electrons reach equipartition with protons and magnetic fields 
after shock acceleration: $\epsel=1/2$. 

 Equations (\ref{syR}) and (\ref{icR}) show that there
are only two parameters left that determine the burst spectrum: $\epsmag$, on
which the spectral peaks depend only weakly, and $\zeta$, the electron injection
fraction, on which the same peaks have a strong dependence. By varying this
injection fraction one can study the dependence on it of the relative intensity 
of the synchrotron and inverse Compton components, and determine those values
of $\zeta$ that maximize the received flux in the BATSE range. Figure 2 shows
the shifting of the burst emission toward higher energies with decreasing
$\zeta$, due to the increase of the electron Lorentz factor. 
For $\zeta > 10^{-2}$ the inverse Compton emission occurs in the Thomson regime 
and carries most of the burst energy, while for $\zeta\siml 10^{-3}$ synchrotron
dominates over inverse Compton scattering, as the latter takes place in the
Klein-Nishina regime. The extent in frequency of 
each component is determined by the ratio $\gamma_M/\gamma_m$ of the maximum
and minimum electron energy of the power-law distribution. The $\gamma_m$ is 
determined by $\epsel/\zeta$ and $\Gamma'_{sh,i/o}$ (\eq [\ref{gmmin}]),
while $\gamma_M$ is set by the details of the electron acceleration and its
calculation is more ambiguous. An upper limit on it is set by requiring
that the radiative cooling timescale is longer than the shock acceleration
timescale, a condition that does not alter the shape of the spectrum near its peak 
unless the injection fraction $\zeta$ is less than $10^{-2}$. 
In all other cases shown in Figure 2, we have set $\gamma_M/\gamma_m=100$, 
which gives the correct shape of the spectrum at frequencies where most
the burst emission lies.

 As shown in Figure 2, there may be more
than one component that carries a good fraction of the burst emission and
each component may extend over a few orders of magnitude in frequency.
This suggests that the radiation which falls in the BATSE window could represent
in some cases a rather small fraction of the entire emission of the burst. 
If we also take into account that at most half of the available internal energy 
can be given to electrons by shock acceleration, it results that the efficiency 
at which the shells' kinetic energy is transformed into radiation visible to 
BATSE can easily be one order of magnitude lower than the efficiency of $\sim 
10\%$ at which the shocks convert the same kinetic energy into internal (for a 
study of the latter, see Kobayashi, Piran \& Sari 1997).  We can draw the
conclusion that the overall process that leads to $\gamma$-ray emission has an 
efficiency  of few percent or lower (see also Daigne \& Mochkovitch 1998, 
Fenimore \etal 1998).
Figure 2 also illustrates the fact that the radiation detected by BATSE can be 
either synchrotron emission or self-inverse Compton scatterings. 
The distinction between the two cases could be made through the detection 
of simultaneous emission at energies well below or above the BATSE range.
 
 Figure 3$a$ shows the spectral evolution of a burst whose BATSE window emission
is due to inverse Compton scatterings. The large low energy ($20-50$ keV) slope 
of the spectra shown in Figure 3$a$ is a result of the monochromaticity 
approximation described in \S3, which was made for numerical reasons. Figure
3$b$ shows the spectral evolution of a synchrotron burst. The shaded curve
represents the average spectrum, calculated as an intensity-weighted average of 
the instantaneous spectra. The curve shown with the thick solid line represents
a fit with the Band function (Band \etal 1993) to the average spectrum, in
the range 30 keV--3 MeV. The fit is characterized by $\alpha=-1.14$ (the
low energy index), $\beta=-2.38$ (the high energy index) and $E_0=326$ keV
($\nu F_{\nu}$ peaks at $(\alpha+2) E_0=282$ keV). 

 As shown in Figure 3$c$
both bursts exhibit a spectral softening that can be well approximated as
a power-law. It should be noted that this is a good fit if the observer would
be able to set $T=0$ when the inner shell is ejected from the burst progenitor.
Obviously, the observer can time the burst only when it begins, in which case 
the power-law spectral softening would be a bad fit. The result shown in 
Figure 3$c$ should be understood as following: there is a $T_0 \sim (1/5-1/3)\;
T_b$, with $T_b$ being the burst duration, such that, if $T$ were measured from 
$T_0$, then the spectral softening would be a power-law in $T$. A clearer
characterization of the burst softening is illustrated in Figure 3$d$, which 
shows the spectral peak decays exponentially with the photon fluence $\Phi_{23}$
in the middle BATSE channels, a feature that was observed in real GRBs by 
Liang \& Kargatis (1996). Within our model, the softening of the burst is due 
to two factors. One is that the shell fluid shocked later is less dense due to 
the radial expansion, leading to lower internal energy densities in the shocked 
gas, which, for a constant parameter $\epsmag$, implies lower magnetic fields. 
Secondly, the radiation emitted by the fluid that moves at larger angles off 
the line of sight toward the center of expansion (the ``central line of sight")
is less blue-shifted by the relativistic motion of the source and arrives later 
at observer than the radiation emitted by the shocked fluid expanding at smaller
angles relative to the central line of sight.
The Lorentz factor of the shocked fluid does not influence the spectral
softening, as it is practically constant during the two-shell interaction.

 In calculating the spectra shown in Figures 3$a$ and 3$b$, we have considered 
the interaction of two shells in which the ejecta are distributed homogeneously.
The general spectral softening is the same if one considers a shell to 
be a collection of ``mini-shells" moving at the same Lorentz factor, \ie if 
the two larger shells have a layered structure. In this case the burst 
light-curve exhibits pulses associated with each new layer that is shocked
(sub-pulse structure). The pulses' fluence depend on the kinetic energies
of the mini-shells, while the separation between pulses is determined by the 
spatial separation between layers. The pulse duration, as well as its shape, 
are determined by the angular extension of the region visible to the observer 
(a spherical cap of half-angle opening $\sim \Gamma_{sh}^{-1}$), the thickness 
of the layer and the electron cooling time-scale. 

The light-curve shown in Figure 4$a$ was obtained considering an inner group 
of layers moving at $\Gamma_i=100$ and an outer one moving at $\Gamma_o=50$. 
The mini-shells in each group have equal masses, corresponding to total shell 
energies $E_i=10^{53}$ ergs and $E_o=2\times 10^{52}$ ergs, if the ejecta are 
spherically symmetric. The energy release parameters (see figure caption) 
where chosen such that the gamma-ray burst is due to synchrotron emission.
Figure 4$a$ shows six individual peaks generated by the shocks that sweep 
six outer layers (in this case the inner mini-shells radiate mostly outside the 
BATSE range), and the 50 keV--300 keV pulse resulting from the addition 
of these pulses. In Figure 4$b$ we show the dependence of the shape of the first 
pulse in the burst shown in Figure 4$a$ on the observing energy. The pulse
lasts longer at lower energy, as observed in real GRBs. If the pulse duration 
is defined by $T_{pulse}^{(i)} = \int_{pulse} (F^{(i)}/F_{max}^{(i)}) dT$, 
where $F^{(i)}$ and $F_{max}^{(i)}$ are the flux and its maximum value in the 
$i$-th observing channel, then we find that $T_{pulse}^{(i)} \propto E_i^{-0.19}$,
where $E_i$ is the geometric mean of the upper and lower energy limits of 
channel $i$. The same dependence is found for the other pulses (Figure 4$c$),
as well as in the case when the radiation in the BATSE range is due to inverse 
Compton emission. The exponent changes slightly if the FWHM of the pulse is used 
instead of the integral duration defined above, if one uses the lower or upper
limits of each BATSE channel instead of their geometric means, or if the photon 
fluxes are used instead of the energetic ones. The dependence found by Norris 
\etal (1996) in real GRBs is $T_{pulse}^{(i)} \propto E_i^{-(0.3 \div 0.4)}$, 
which is stronger than shown by our numerical bursts.

 The usual argument used to explain the observed pulse duration -- energy 
anti-correlation is that the electron synchrotron cooling timescale 
$t_{sy} \propto \gamma_e^{-1} \propto (h\nu_{sy})^{-1/2}$. 
Because the synchrotron spectrum of an electron extends over more than just 
one BATSE channel, the above exponent of $1/2$ should be regarded only as an 
upper limit, which is consistent with observations. However the argument based
on the electron cooling ignores the possible contribution of the geometrical 
curvature of the layer and of its thickness to the pulse duration.  
The spread in photon arrival time due to the curvature of the shell is 
$T_{\theta} \sim t/(2\Gamma_{sh}^2)$, where $t$ is the lab-frame time, 
because the observer receives radiation mainly from the fluid moving within 
$\theta \sim \Gamma_{sh}^{-1}$ off the central line of sight. If the observer 
frame electron cooling timescale $T_{sy} \sim t_{sy}/(2\Gamma_{sh}^2)$ exceeds 
$T_{\theta}$, then the shell radius $r=ct$ and its volume increase significantly 
during $t_{sy}$, leading to excessive adiabatic losses 
and to a lower burst efficiency. Thus, an efficient burst is one where 
$T_{sy} < T_{\theta}$. Our choice of model parameters ensures that the electrons 
are radiative, and implies that the pulse duration -- energy anti-correlation
is not due to the electron cooling\footnotemark.
\footnotetext{Because the sweeping up of a 
layer is simulated through the addition of a single cell of shocked fluid 
(which is done for numerical reasons), it follows that the duration dependence 
with energy shown by the numerically simulated pulses is due only to the geometrical 
curvature of the emitting shell}

 In Figure 4$d$ we show the six pulses that form the burst shown in Figure 4$a$, 
as seen in the 50 keV--300 keV band. The fluxes have been normalized to their maxima
and the pulses have been aligned at their peaks. It can be noticed that these 
pulses have a sharp rise and a slow decay and that are more asymmetric than 
the average pulse shape determined by Norris \etal (1996) for separable pulses 
in long and bright GRBs. The pulse shape is determined by the relative 
importance of the layer thickness, its angular spreading and the electron 
cooling. As mentioned before, for our choice of energy release parameters the 
electron cooling is too fast to play any part in determining the pulse shape.
If the thickness of the layer is not taken into account, then the emitting
region is approximated by a surface and the emission of radiation is 
almost instantaneous. This explains the similarity between the shape
of the numerical pulses and that obtained analytically by Fenimore, Madras
\& Nayakshin (1996) for the pulse resulting from the interaction between a 
single shell and a stationary medium, when the shell is approximated as 
infinitesimally thin and the radiated power as a delta-function in time.
The effect of taking into account the shell thickness can be assessed
by comparing the light-curves shown in Figures 4$a$ and 4$b$, corresponding
to a shell containing several thin layers and one infinitesimally thin
layer, respectively. Thus, a shell thickness $\sim r/\Gamma^2$ yields a slightly
more symmetric pulse, and a shell thickness larger than predicted by 
\Mesz \etal (1993) is required to obtained an even more symmetric pulse. 

 The above equation for the spread in the photon arrival time due to the
geometric curvature of the source ($T_{\theta} \sim t/(2\Gamma_{sh}^2)$) 
implies a correlation between the pulse duration and pulse onset time for
all the pulses arising from the collision of two inhomogeneous shells. 
This correlation is illustrated in Figure 4$d$, which clearly shows that 
later pulses last longer than earlier ones. However, such a correlation 
will not be manifested by all the pulses in a real GRB, as it is possible 
that pulses seen close to each other by the observer were emitted by 
different groups of colliding mini-shells, located at different radii, 
and thus having different durations. This is illustrated in Figure 5, where 
we considered 10 pairs of interacting shells. Each pair has the same parameters 
as the two shells that yield the burst shown in Figure 4$a$, except that the 
time interval $t_v$ between their ejection differs from pair to pair. 
Therefore each pair has a different interaction radius. The time elapsed 
between the ejection of successive pairs is also considered variable. 
The light-curve shown in the Figure 5$a$ was calculated assuming that, in 
the frame of the shocked fluid, the emission is concentrated within two cones
of solid angle $4\pi/5$ sr directed outward and inward along the radial 
direction, so that the observer receives radiation from the 
shocked gas moving only within $\sim \Gamma_{sh}^{-1}/2$ off the central 
line of sight. This was done in order to reduce the spread in the photon
arrival time due to the angular spread of the shell and the pulse overlapping
which is present in Figure 4$a$. A random injection of several groups of shells
can be simulated by repeating and superposing the template pulse shown in 
Figure 5$a$, corresponding to arbitrary values of the pulse onset time, 
intensity and duration determined by the value of $r_{int}$ for each pair of 
interacting shells. We chose for simplicity a periodic shell Lorentz factor 
and mass distribution in the wind which ensures that each shell suffers only 
one collision. An example of such a complex light-curve simulation is shown 
in Figure 5$b$.

\vspace*{3mm}
\section{Conclusions}

As illustrated by the numerical results shown in the previous section,  the
observed GRB $\gamma$-rays can be due either to the synchrotron emission from  
shocked ejecta, if the electron injection fraction is small enough 
(typically around $10^{-2}$) to ensure that the accelerated electrons 
reach high random Lorentz factors, or to the up-scattering of the 
synchrotron photons, if the electron injection fraction is not far below 
unity. The particular choice of the shell Lorentz factors used in this work 
(determined by numerical reasons) has lead to an overall efficiency 
of converting the initial shell kinetic energy into $\gamma$-rays in the 
range 25 keV--1 MeV that is below 1\%. A wider range of Lorentz factors can 
increase this efficiency. Nevertheless, since only a fraction of the internal 
energy of the shocked gas can be given to the electrons through shock 
acceleration, and the spectral emission range of the burst is very broad, 
the efficiency in a given instrument's range (such as BATSE) can be 
substantially smaller than that calculated using only the dynamics of the 
interaction between shells. 

The fact that the emission from the gas moving at larger angles relative to 
the central line of sight arrives later to the observer, and has 
a softer spectrum than the emission from the fluid flowing at smaller 
angles, leads to an increase in the pulse duration with decreasing energy. 
We find that, if the geometrical curvature of the shell were the only 
factor that determines the pulse duration, then the pulse duration 
dependence on energy would be $E^{-0.19}$, which is weaker than observed. 
The pulse duration is also determined by the electron cooling timescale 
and by the shell thickness. This inconsistency with observations may be 
due to our choice in the numerical calculations of parameters that led to
an electron cooling timescale smaller than that of the adiabatic losses, to 
maximize the burst efficiency, and to an interaction radius sufficiently large 
to ensure optical thinness, which led to a spread in the photon arrival 
time due to shell's curvature dominating that due to its thickness.  

A softening of the burst spectrum with time is a natural consequence of the
above-mentioned correlation between the angle relative to the observer
at which the emitting fluid moves, the arrival time and the hardness 
of the radiation received, together with the progressive decrease of the 
turbulent magnetic field intensity due to the radial expansion of the ejecta.
The evolution of the break energy of the numerical spectra can be approximated 
quite well as an exponential in the 50 keV--300 keV photon fluence,
for bursts in which either the synchrotron or the inverse Compton emission 
peaks around 100 keV. The spectrum of the synchrotron burst is well 
approximated by the Band function.

As shown by Daigne \& Mochkovitch (1997), a significant subset of the 
spectral-temporal correlations observed can be explained within a simple 
treatment of the kinematics and dynamics of unsteady winds.  Here we have 
shown that in a more complete radiation and hydrodynamical treatment, some 
of the burst features found by Daigne \& Mochkovitch (1997) are qualitatively 
confirmed, while some quantitative details differ, possibly because of the 
more detailed physics (including the hydrodynamic treatment) introduced 
here. In particular, our treatment is able to reproduce an observational
feature that previously was not easily obtained, namely a spectral softening 
in time. 

 The ``kinematic" light-curves calculated by Kobayashi \etal (1997) for 
internal shocks exhibit a complicated structure, and their bolometric temporal 
profiles bear a good resemblance to those of real GRBs. Our band light-curves
show a similar behavior, and can in addition probe the physical origin of 
more detailed effects, such as spectral--temporal correlations. In the case 
where all pulses within a burst arise from one single group of interacting shells, 
one would expect a correlation between the pulse duration and the time measured 
from the beginning of the burst. This is due to the fact that, on average, 
successive collisions within a single group of shells take place at larger radii, 
and that the pulse onset time and duration are both proportional with the radius 
where each collision takes place. Obviously, the pulses seen in a real GRB may be 
due to several groups of shells interacting at different radii, and thus 
producing sets of pulses of different durations that overlap and mix, as 
illustrated in Figure 5b. 
 
 Whereas a pulse duration increase with time is the signature of the 
ejection of closely bunched shells with different Lorentz factors, a lack of 
continuous correlation between pulse duration and pulse onset time would 
indicate repeated episodes, stretching over a longer period of time, of 
ejection of bunches of shells. This may be useful in mapping the injection
time-history by the central engine, and perhaps shed some light on the 
dynamics of the post-collapse or merger-disruption event. For instance,
the above correlation (or lack thereof) could be used for testing
whether double (or multiple) peaked bursts arise from discrete and separated
``events". Examples of such discrete events might be, \eg, the accretion 
of discrete rings of disrupted matter (or more speculatively, the collapse to a 
neutron star followed by collapse to a black hole or the collapse of a primary 
followed by explosive deleptonization of a small mass neutron star companion). 
Each discrete event would be characterized by the above correlation within
the event, which then resets itself at the next event, if they are truly 
discrete and independent.

 The results published so far on the GRBs produced by internal shocks 
in an unsteady relativistic wind show that this model is able to explain
many of the well established properties and correlations observed 
in real bursts. Further work is necessary to analyze the model features at a 
more detailed level: numerical results having sufficient temporal resolution 
would allow a comparison with the other correlations among pulse features 
(peakedness, asymmetry, width, centroid lag) found by Norris \etal (1996). 
Such features, as well as the general efficiency, are issues that may need 
to be addressed within more specific models for the burst progenitor. The
advantage of the calculations presented here is that they are independent of 
any specific model about the primary event, the only requirement being that 
the central engine produces a sufficiently energetic relativistic wind.

\acknowledgements{This research was supported in part by NASA NAG5-3801 and
NAG5-2857}

\onecolumn

\begin{appendix}

\vspace*{5mm}
\section{Relativistic Shocks: $\Gamma_i \gg \Gamma_o$}

\vspace*{3mm}
 If the shocks are relativistic ($G \gg 1$), the solution of equation
(\ref{xeq}) leads to
\begin{equation}
 \Gamma'_{sh,i} \sim \frac{1}{2} \left(\frac{G^3}{(G-y)(Gy-1)}\right)^{1/2}, \quad
 \Gamma_{sh} = \left(\Gamma_{i}\Gamma_o\frac{Gy-1}{G-y}\right)^{1/2}, \quad
 \Gamma'_{sh,o} = y \Gamma'_{sh,i}\;,
\label{GprimR}
\end{equation}
provided that $2/G < y < G/2$ (otherwise the shock
propagating in the denser shell cannot be considered relativistic).
From equation (\ref{P}) and assuming the adiabatic index $\hat{\gamma}=4/3$ 
for a hot gas, the internal energy density in the shocked fluid is
\begin{equation}
 e = \sqrt{\rho_i\rho_o} c^2  \left[\frac{G^3y}{(G-y)(Gy-1)}\right]\;,
\end{equation} 
and, using equations (\ref{rint}) and (\ref{rho}) to derive the pre-shock
co-moving densities, the magnetic field can be calculated:
\begin{equation}
 B = 9.5 \times 10^3 \,\epsmg1^{1/2} E_{i,53}^{1/2}
     \Gamma_{o,2}^{-3} t_{v,0}^{-3/2}
    \left[\frac{G^3}{(G-y)(Gy-1)}\right]^{1/2}\; [{\rm G}] \;,
\end{equation}
where $\epsmag$ is a parameter describing the magnetic field strength (see \S3) 
and where the usual notation $A=10^nA_n$ was used.
The minimum electron Lorentz factor $\gamma_m$ of the power-law distribution 
of shock accelerated electrons is 
\begin{equation}
 \gamma_{m,i/o} \sim \frac{1}{3} \frac{m_p}{m_e} \frac{\epsel}{\zeta}
  (\Gamma'_{sh,i/o}-1) \;,
\label{gmmin}
\end{equation}
where $\epsel$ and $\zeta$ parameterize $\gamma_m$ and an electron index $p=2.5$ 
was assumed (see \S3). Together with equation (\ref{GprimR}), equation 
(\ref{gmmin}) leads to
\begin{equation}
 \gamma_{m,i} \sim 300\, \frac{\epsel}{\zeta}
   \left[\frac{G^3}{(G-y)(Gy-1)}\right]^{1/2} = \frac{\gamma_{m,o}}{y} \;.
\end{equation}

 The observed peaks of the synchrotron and inverse Compton emission are 
straightforward to calculate:
\begin{equation}
 h\nu_{sy,i} = 1.6\, \left(\frac{\epsel}{\zeta}\right)^2
     \epsmg1^{1/2} E_{i,53}^{1/2} \Gamma_{o,2}^{-2} t_{v,0}^{-3/2}
     \left[\frac{G^5}{(G-y)^2(Gy-1)}\right]\; [{\rm keV}] \; =
     \frac{h\nu_{sy,o}}{Y} \;,
\label{syR}
\end{equation}
\begin{equation}
   h\nu_{ic,i} = 140\, \left(\frac{\epsel}{\zeta}\right)^4
     \epsmg1^{1/2} E_{i,53}^{1/2} \Gamma_{o,2}^{-2} t_{v,0}^{-3/2}
      \left[\frac{G^8}{(G-y)^3(Gy-1)^2}\right]\; [{\rm MeV}] \; =
     \frac{h\nu_{sy,o}}{Y^2} \;,
\label{icR}
\end{equation}
assuming that the up-scattering of photons takes place in the Thomson regime.

 The lab-frame electron radiative cooling timescale is upper bounded by the 
synchrotron cooling time
\begin{equation}
 t_{sy,i} = 2.9\, \left(\frac{\epsel}{\zeta}\right)^{-1} 
     \epsmg1^{-1} E_{i,53}^{-1} \Gamma_{o,2}^{7} t_{v,0}^{3} 
    \left[\frac{(G-y)(Gy-1)^2}{G^4}\right]  \; [{\rm s}] \; = y t_{sy,o} \;,
\label{tsy}
\end{equation}    
which is much shorter than the timescale for adiabatic losses $t_{ad} \sim
t_{int}$, where, from equation (\ref{rint}), $t_{int} = r_{int}/c \sim 2 
\times 10^4\, \Gamma_{o,2}^2 t_{v,0}$ s. 
Equation (\eq [\ref{tsy}]) shows that for model parameters that
are not far from the chosen scaling values, the synchrotron cooling timescale 
of the electrons radiating at $h\nu \sim 100$ keV is smaller by four 
orders of magnitude than the time during which the shock sweeps up the 
shell, which is of order $t_{int}$. For the observer, the electron cooling 
time appears $\sim \Gamma_{sh}^2$ times shorter, due to the motion of the 
source, and is therefore much smaller than the spread in the photon arrival
time due to the shell curvature, which is of order $t_{int}/\Gamma_{sh}^2 \sim t_v$. 

\clearpage

\vspace*{5mm}
\section{Synchrotron and Inverse-Compton Emission}

\vspace*{3mm}
{\bf Synchrotron emission.}
The calculation of the synchrotron spectrum is based on a numerical approximation 
derived from the equations given by Rybicki \& Lightman (1979). The synchrotron 
power per unit frequency $P(\omega)$ (for one electron) is
\begin{equation}
 P(\omega) = \frac{3^{5/2}}{8\pi} \frac{P_{sy}}{\omega_c} 
             F\left(\frac{\omega}{\omega_c}\right)\,; 
\end{equation}
where $P_{sy}=(1/6\pi) \sigma_{Th} c B^2 (\gamma_e^2-1)$ is the frequency-integrated 
synchrotron power (and averaged over the pitch angle), $\sigma_{Th}$ being the 
cross-section for electron scattering and $\gamma_e$ the electron Lorentz factor, 
and $\omega_c=(3\pi/8) (eB/m_e c) \gamma_e^2$ is the synchrotron frequency 
(averaged over the pitch angle), with $e$ and $m_e$ the electron charge and mass, 
respectively. $F(u)$ can be approximated by  $F(u) \sim 1.78\, u^{0.297}e^{-u}$
for $10^{-3.5} < u < 10^{0.5}$ with a maximum error of 5\%. For $u$ such that 
$F(u) > 0.5$ (\ie close to the peak of the synchrotron spectrum) this approximation 
is accurate to better than 1\%.  At frequencies far from the synchrotron peak, we used
 the approximations given by Rybicki \& Lightman (1979): $F(u) \sim 2.15\, u^{1/3}$ 
for $u \ll 1$ and $F(u) \sim 1.25\, e^{-u} \sqrt{u}$ for $u \gg 1$.

\vspace*{3mm}
 {\bf Inverse Compton scatterings.}
The peak of the inverse Compton spectrum and the inverse Compton losses are 
calculated using approximations derived from the equations given by Blumenthal 
\& Gould (1970). The inverse Compton spectrum peaks at the energy $\epsilon_p$ given by
\begin{equation}
 \epsilon_p = \frac{4 q_p \gamma_e^2 \epsilon_0}
              {1+(4 q_p \gamma_e \epsilon_0/m_e c^2)} \;,
\label{epsp}
\end{equation}
where $\epsilon_0$ is the energy of the incident photon, and $q_p$ is a factor that
depends weakly on $\gamma_e \epsilon_0$. We found that $q_p$ can be approximated with 
an error below 1\% by
\begin{equation}
 q_p = \frac{1}{2} + \frac{5.91 \times 10^{-2}}{1+0.184 v^{1.31}}
                   + \frac{5.09 \times 10^{-2}}{1+51.6 v^{1.45}}  \,;
\label{qp}
\end{equation}
with $v \equiv \gamma_e \epsilon_0 /m_e c^2$. Equations (\ref{epsp}) and (\ref{qp})
lead to $q_p=0.610$ and $\epsilon_p \sim 2.44 \gamma_e^2 \epsilon_0$ in the
Thomson regime ($v \ll 1$), and to $q_p=1/2$ and $\epsilon_p \sim \gamma_e m_e c^2$
in the extreme Klein-Nishina regime ($v \gg 1$). For the inverse Compton power
$P_{ic}$ (per electron) we found that the following approximation: 
\begin{equation}
 \frac{P_{ic}(v)}{P_{ic}^{Th}} \sim \left\{ \begin{array}{ll}
     [1+7.67 \exp (2.43 \log v)]^{-1} & v \leq 1 \\
     0.107 v^{-1.07} \exp (-0.569 \log^2 v) & 1 < v < 10^{1.5} \end{array} \right. \;,
\end{equation}
where $P_{ic}^{Th} = (4/3) \sigma_{Th} c U_{sy}^2 (\gamma_e^2-1)$ is the inverse
Compton power in the Thomson regime, and $U_{sy}$ the energy density of the
photon field that is up-scattered, has a relative error that increases with $v$,
reaching a maximum value of 10\% at $v=10^{1.5}$, where $P_{ic} \sim 10^{-3} P_{ic}^{Th}$.
The inverse Compton losses are severely reduced by the Klein-Nishina effect at
$v > 10^{1.5}$ and an accurate treatment of these losses is not necessary.

\end{appendix}


\input PSfig

\clearpage

\begin{figure}
\centerline{\psfig{figure=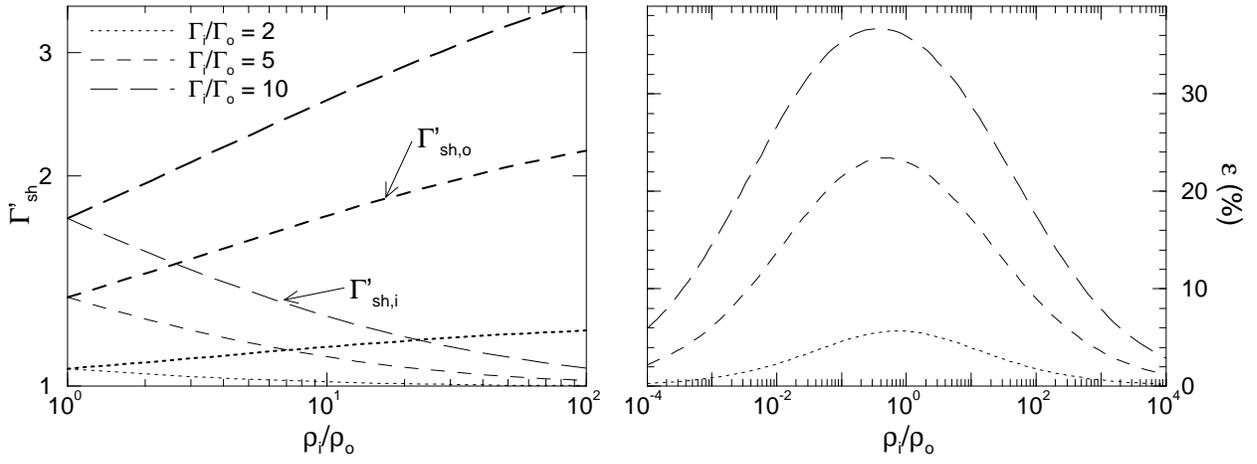}}
\vspace*{2cm}
\caption{Left panel: Lorentz factors of the shocked fluid as measured in the 
 frames of the yet un-shocked gas in the inner shell ($\Gamma'_{sh,i}$ -- 
 thin lines) and in the outer one ( $\Gamma'_{sh,o}$ -- thick lines), for a 
 range of shells' density ratio and for three values of the ratio of their 
 lab-frame Lorentz factors. For given ratio $\Gamma_i/\Gamma_o$, the Lorentz 
 factors $\Gamma'_{sh,i}$ and $\Gamma'_{sh,o}$ are symmetric relative to the 
 ordinate. Right panel: the efficiency of the shocks in converting the shells' 
 total kinetic energy into internal energy. The legend is the same as for the 
 left panel.}
\end{figure}

\begin{figure}
\vspace*{-6cm}
\centerline{\psfig{figure=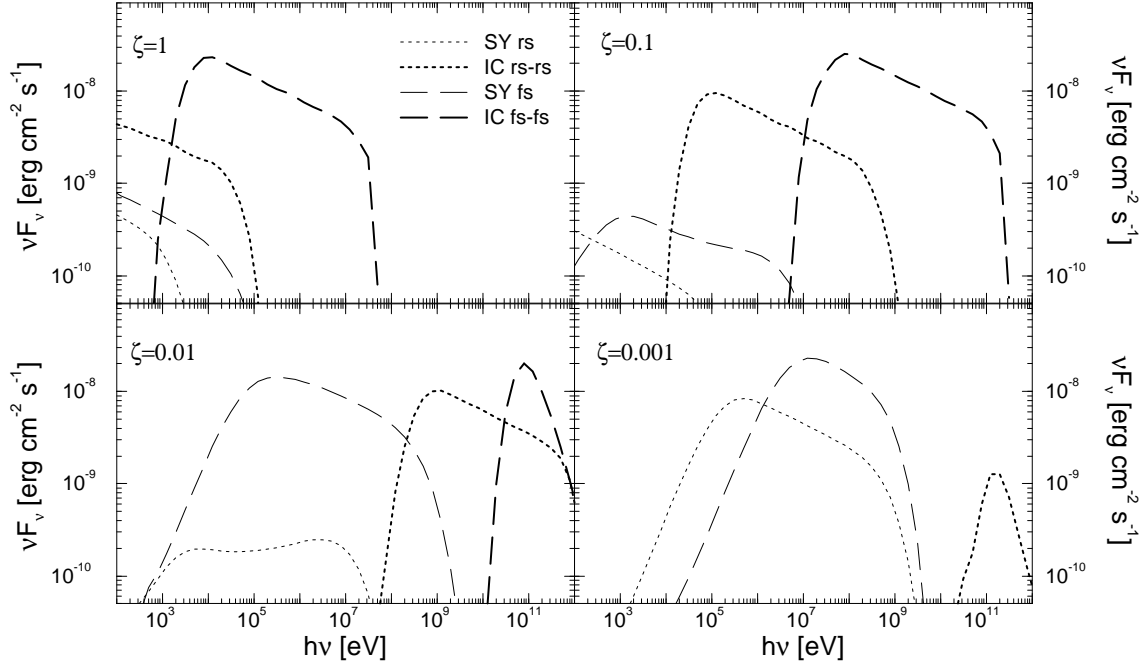}}
\vspace*{-1cm}
\caption{Dependence of burst spectra on $\zeta$, the electron injection 
 fraction, for fixed $E_i=10^{53}$ ergs and $E_o= 2 \times 10^{52}$
 ergs in $4\pi$ sr, $\Gamma_i=100$, $\Gamma_o=50$, and $t_v=1$ s, 
 $\epsel=1/2$, $\epsmag=10^{-1}$ and $p=2.5$. The source is located at redshift 
 $z=1$, with $H_0=75\,{\rm km\,s^{-1} Mpc^{-1}}$ and $\Omega = 1$. The synchrotron 
 (thin lines) and self-inverse Compton (thick lines) emissions from the reverse 
 shock (rs) and forward shock (fs) are shown separately. The mixed components
 resulting from up-scattering by one of the shocks of synchrotron photons 
 generated by the other shock are too weak and do not appear in the graphs.
 Note that for $\zeta \simg 10^{-1}$, the observer receives up-scattered photons 
 in the BATSE range, while for $\zeta \siml 10^{-2}$ the $\gamma$-ray burst is 
 due to synchrotron emission, and the inverse Compton emission is diminished
 by the Klein-Nishina reduction.}
\end{figure}

\begin{figure}
\vspace*{1cm}
\centerline{\psfig{figure=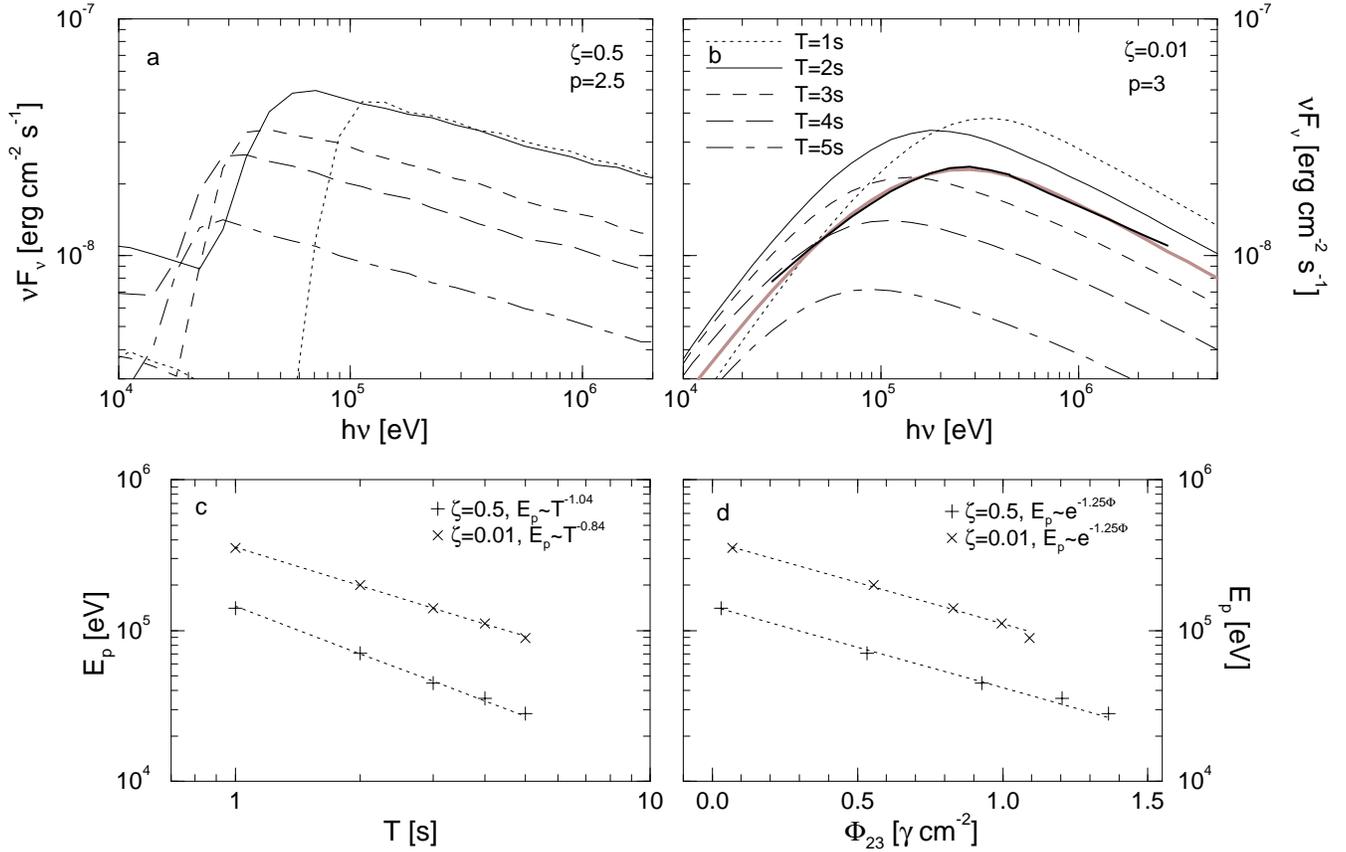}}
\vspace*{2cm}
\caption{Spectral evolution of bursts with same parameters as in 
 Figure 2, except $\zeta=0.5$ for panel $a$ (inverse Compton emission)
 and $p=3$ for the panel $b$ (synchrotron emission). The average spectrum
 shown in graph $b$ with a shaded continuous line is fit in the range
 30 keV--3 MeV by the Band function with parameters $\alpha=-1.14$,
 $\beta=-2.38$ and $E_0=326$ keV. 
 Graph $c$ shows in log-log scale the decrease with time of $E_p$, the peak of 
 $\nu F_{\nu}$ (power-per-decade), \ie the softening of the burst spectrum.
 x's are for synchrotron emission and pluses for inverse Compton radiation 
 in BATSE window. For both types of bursts, a power-law approximates quite 
 well the spectral softening. Graph $d$ shows the exponential decay of the 
 peak $E_p$ with the photon flux $\Phi_{23}$ in the range 50 keV--300 keV.} 
\end{figure}

\begin{figure}
\centerline{\psfig{figure=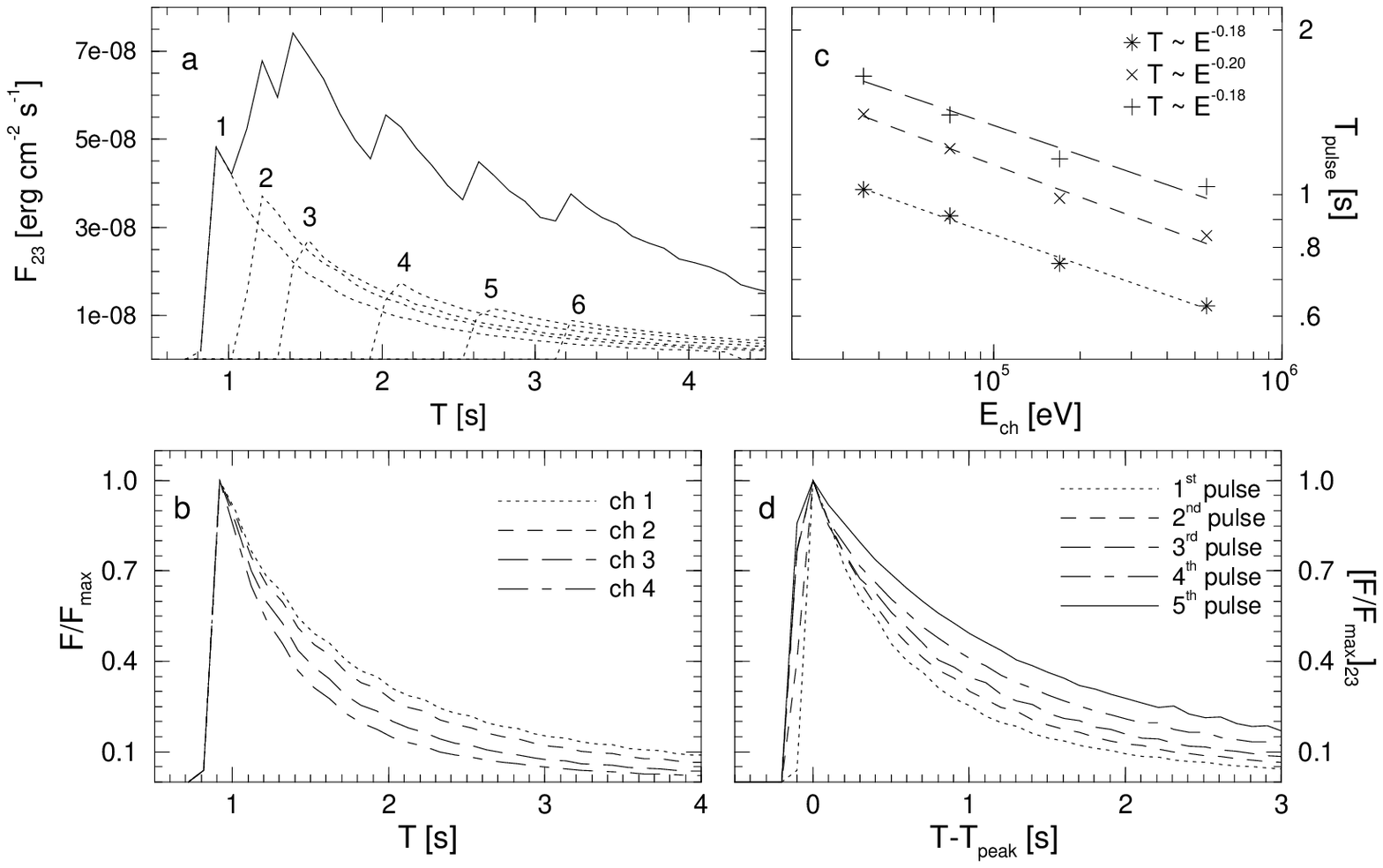}}
\caption{\newline $a$ : Light-curve of a burst arising from the collision between
 an inner set of faster and more massive shells with a group of six outer, slower 
 and less massive shells. The Lorentz factor and total kinetic energy of the 
 shells is each group are $\Gamma_i=100, E_i=10^{53}$ ergs (assuming spherical
 symmetry), and $\Gamma_o=50, E_o=2 \times 10^{52}$ ergs. The burst is
 located at redshift $z=1$. Other parameters are  $\epsel=1/2$, $\zeta=10^{-2}$,
 $\epsmag=10^{-1}$. \newline
 $b$ : The first pulse in graph $a$, as seen in each BATSE channel. 
       The pulse lasts longer at lower energies.  \newline
 $c$ : The dependence on observing energy of the duration of the first three
      pulses shown in graph $a$. \newline 
 $d$ : Evolution of the pulse duration with pulse onset time. For a collision 
      between two layered shells, later pulses last longer than earlier ones.}
\end{figure}

\begin{figure}
\centerline{\psfig{figure=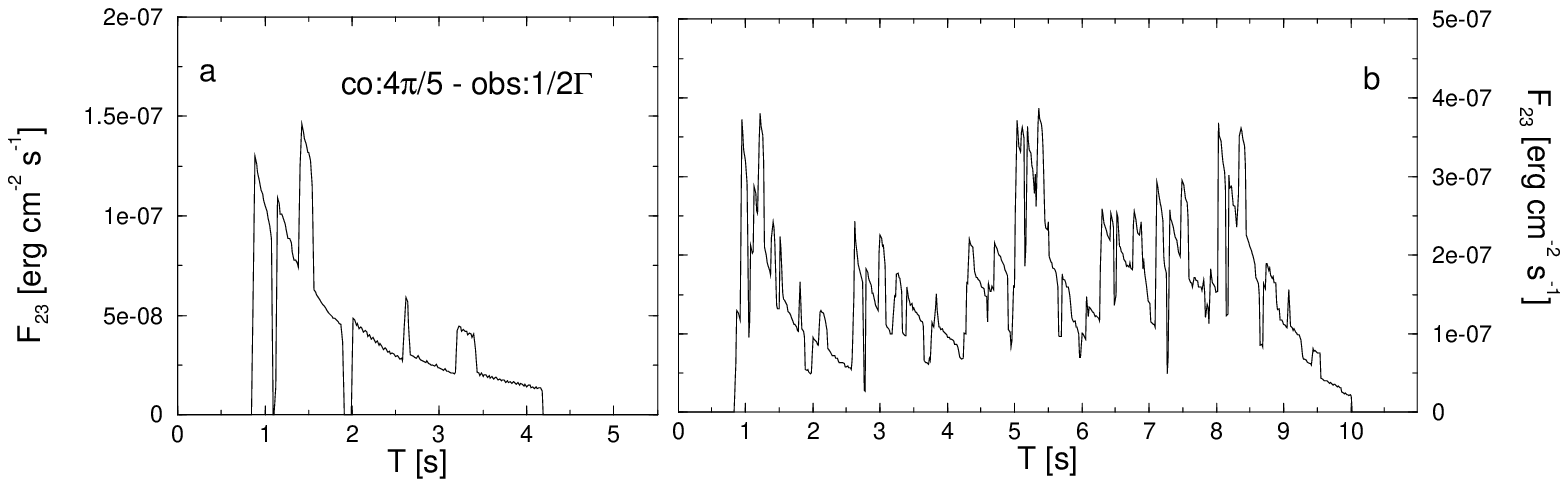}}
\caption{Light-curves arising from pairs of shells having the same
  parameters as the pair that yields the burst shown in Figure 4$a$,
  for a pre-beaming factor $4\pi/5$ of the comoving radiation. 
  Panel $a$ shows the pulse from a single pair, while panel $b$ shows the 
  light-curve from a set of 10 pairs (see text for details).}
\end{figure}


\begin{references}

 Band, D., \etal 1993, ApJ, 413, 281                                           \\
 Bhat, P.N., \etal 1994, ApJ, 426, 604                                         \\
 Blandford, R.D. \& McKee, C.F. 1976, Phys. Fluids, 19, 1130                  \\
 Blumenthal, G.R. \& Gould, R.J. 1970, Rev. Mod. Phys., vol. 42, no. 2, 237  \\
 Daigne, F. \& Mochkovitch, R. 1998, MNRAS, 296, 275                        \\
 Fenimore, E.E., \etal 1998, ApJ, submitted (astro-ph/9802200)             \\
 Fenimore, E.E., Madras, C. D., \& Nayakshin, S. 1996, ApJ, 473, 998      \\
 Ford, L.A., \etal 1995, ApJ, 428, 620                                   \\
 Kobayashi, S., Piran, T. \& Sari, R. 1997, ApJ, 490, 92                \\
 Kouveliotou, C., \etal 1993, ApJ, 413, L101                           \\
 Liang, E. \& Kargatis, V. 1996, Nature, 381, 49                      \\
 \Mesz, P., Laguna, P. \& Rees, M.J. 1993, ApJ, 415, 181             \\
 Norris, J.P., \etal 1996, ApJ, 459, 393                            \\
 Panaitescu, A. \& \Mesz, P. 1998a, ApJ, 492, 683                  \\
 Panaitescu, A. \& \Mesz, P. 1998b, ApJ, 501, 772                 \\
 Papathanassiou, H. \& \Mesz, P. 1996, ApJ, 471, L91             \\
 Papathanassiou, H. \& \Mesz, P. 1998, in preparation           \\
 Pilla, R. \& Loeb, A. 1998, ApJ, 494, L167                    \\
 Rees, M.J. \& \Mesz, P. 1994, ApJ, 430, L93                  \\
 Rybicki, G.B. \& Lightman, A.P. 1979, {\sl Radiative Processes in
           Astrophysics} (New York:Wiley-Interscience)            \\
 Sari, R., \& Piran, T. 1997, MNRAS, 287, 110                \\
 Wen, L., Panaitescu, A. \& Laguna, P. 1997, ApJ, 486, 919     \\

\end{references}
\end{document}